\documentclass[conference]{IEEEtran}
\IEEEoverridecommandlockouts
\usepackage{cite}
\usepackage{amsmath,amssymb,amsfonts}
\usepackage{algorithmic}
\usepackage{graphicx}
\usepackage{textcomp}
\usepackage{tcolorbox}
\usepackage{xcolor}
\usepackage{array}
\usepackage{tikz}
\usepackage{adjustbox}
\usepackage{fontawesome5}
\usepackage{url} 
\usepackage{hyperref}
\usepackage[edges]{forest}
\tikzset{%
    parent/.style =          {align=center,text width=3.5cm,rounded corners=3pt},
    child/.style =           {align=center,text width=3.5cm,rounded corners=3pt},
    grandchild/.style =      {align=center,text width=3.5cm,rounded corners=3pt},
    greatgrandchild/.style = {align=center,text width=1.5cm,rounded corners=3pt},
    referenceblock/.style =  {align=center,text width=1.5cm,rounded corners=2pt}
}
\usepackage{float}
\usepackage{placeins}

\usepackage{comment}
\def\BibTeX{{\rm B\kern-.05em{\sc i\kern-.025em b}\kern-.08em
    T\kern-.1667em\lower.7ex\hbox{E}\kern-.125emX}}

\newtcolorbox{chatbox}{
  colback=gray!5!white, colframe=white!60!black,
  boxrule=0.5pt, arc=0mm, boxsep=2pt,
  left=6pt,right=6pt,top=4pt,bottom=4pt,
  width=\columnwidth,
  fontupper=\footnotesize\color{black}  
}

\tcbset{
  mybox/.style={
    enhanced,
    sharp corners,
    boxrule=0.5pt,
    colback=gray!5!white,
    width=\columnwidth,
    left=4pt,right=4pt,top=4pt,bottom=4pt,
    halign=flush left,
    fontupper=\footnotesize\color{black}
  }
}

\begin{document}

\title{Enhancing Clinical Decision Support and EHR Insights through LLMs and the Model Context Protocol: An Open-Source MCP-FHIR Framework}

\author{
\IEEEauthorblockN{ Abul Ehtesham\textsuperscript{1},   Aditi Singh\textsuperscript{2}, Saket Kumar\textsuperscript{3}}
\IEEEauthorblockA{
\textsuperscript{1}\textit{Kent State University, USA} \\
\textsuperscript{2}\textit{Department of Computer Science, Cleveland State University, USA} \\\textsuperscript{3}\textit{Northeastern University, USA}\\
\texttt{aehtesha@kent.edu, a.singh22@csuohio.edu, kumar.sak@northeastern.edu}
}
}
\maketitle
\begin{abstract}
Enhancing clinical decision support (CDS), reducing documentation burdens, and improving patient health literacy remain persistent challenges in digital health. This paper presents an open-source, agent-based framework that integrates Large Language Models (LLMs) with HL7 FHIR data via the Model Context Protocol (MCP) for dynamic extraction and reasoning over electronic health records (EHRs). Built on the established MCP-FHIR implementation, the framework enables declarative access to diverse FHIR resources through JSON-based configurations, supporting real-time summarization, interpretation, and personalized communication across multiple user personas, including clinicians, caregivers, and patients. To ensure privacy and reproducibility, the framework is evaluated using synthetic EHR data from the SMART Health IT sandbox (https://r4.smarthealthit.org/), which conforms to the FHIR R4 standard. Unlike traditional approaches that rely on hardcoded retrieval and static workflows, the proposed method delivers scalable, explainable, and interoperable AI-powered EHR applications. The agentic architecture further supports multiple FHIR formats, laying a robust foundation for advancing personalized digital health solutions.
\end{abstract}


\section{Introduction}
Despite the widespread adoption of electronic health records (EHRs)~\cite{CMS_EHR_Overview}, significant challenges persist in enhancing clinical decision support (CDS), reducing physician documentation burdens, and improving patient comprehension of health information. While the integration of HL7 Fast Healthcare Interoperability Resources (FHIR)~\cite{HL7FHIROverview} and mandates such as the 21st Century Cures Act~\cite{21stcentury} have expanded access to structured medical records, a substantial gap remains between data availability and its meaningful interpretation by end users particularly clinicians, caregivers, and patients~\cite{graham2008}.

Recent advances in large language models (LLMs)~\cite{zhao2025surveylargelanguagemodels}, including OpenAI's GPT-4~\cite{openai_gpt4} and emerging open-source alternatives~\cite{minaee2025largelanguagemodelssurvey}, offer promising capabilities in summarizing, interpreting, and simplifying complex medical content. However, integrating LLMs into clinical workflows~\cite{Davis2019ClinicalWorkflows} continues to face obstacles related to effective context injection, consistent access to structured EHR data, and ongoing concerns about replicability, safety, and explainability in medical AI systems~\cite{nori2023capabilities,teo2023notready}.

\begin{figure*}
    \centering
    \includegraphics[width=0.9\linewidth]{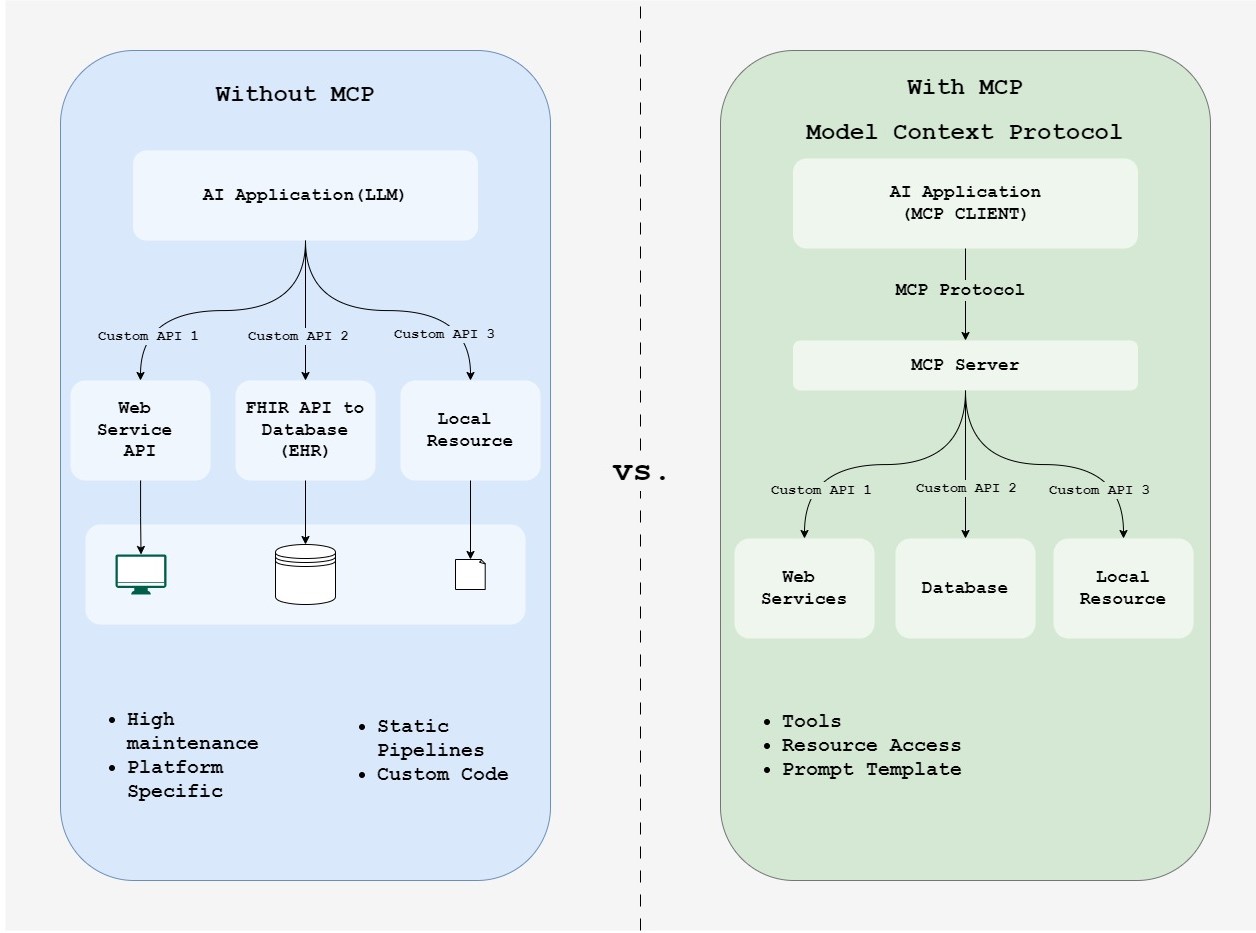}
    \caption{Comparison of traditional AI application integration (left), where the LLM directly calls multiple custom APIs leading to high maintenance and platform-specific pipelines—versus the Model Context Protocol (MCP) approach (right), which centralizes access to tools, resources, and prompt templates through a unified MCP server.}
    \label{fig:comparison}
\end{figure*}

Earlier efforts, such as the LLM on FHIR system~\cite{schmiedmayer2024llmfhirdemystifying}, demonstrated that mobile applications could retrieve and interpret patient records by leveraging LLMs and function-calling mechanisms built on Stanford's Spezi ecosystem~\cite{StanfordSpezi2025}. While these solutions showed the feasibility of converting structured medical data into user-friendly narratives, their broader clinical adoption has been limited due to platform-specific constraints, static data pipelines, and inconsistent LLM outputs.
To address these limitations, this paper presents an open-source, agent-based framework that integrates Large Language Models with the Model Context Protocol (MCP)~\cite{202504.0245} and a dynamic FHIR server, using the publicly available MCP-FHIR implementation~\cite{mcpfhir}. Through declarative JSON configurations and RESTful interactions, the system enables LLMs to retrieve and summarize diverse FHIR-based patient records without the need for custom hardcoding. This modular design facilitates transparent, reproducible reasoning workflows across heterogeneous EHR systems and varying FHIR formats. Figure ~\ref{fig:comparison} illustrates the architectural contrast between traditional, manually wired integrations and the MCP-based approach. While the former relies on hardcoded API calls and static pipelines, MCP enables dynamic, declarative tool invocation and composable agent workflows.

The framework also incorporates the MCP-Agent module~\cite{mcpagent2025}, which abstracts server interactions and enables dynamic orchestration of AI agents. By supporting composable design patterns, MCP-Agent simplifies the development of robust, model-agnostic agents and enhances the flexibility and interoperability of the system.

The remainder of this paper is structured as follows. Section~II reviews related work on LLM integration in clinical applications. Section~III details the system architecture, including the roles of the MCP server, LLM engine, and patient-facing interface. Section~IV describes the implementation workflow. Section~V presents a practical use case demonstration. Finally, Section~VI concludes the paper.

\section{Related Work}
The rapid advancement of large language models (LLMs) in recent years has catalyzed a wide range of healthcare applications~\cite{he2025surveylargelanguagemodels}, spanning clinical decision support, documentation, and patient education. Early research demonstrated the potential of LLMs to assist clinicians with documentation and summarization tasks~\cite{shah2023creation,nori2023capabilities}. These studies showed that LLMs could extract meaningful insights from unstructured text; however, reliably incorporating structured electronic health record (EHR) data into clinical workflows remained a significant challenge.

Initiatives such as the LLM on FHIR system~\cite{llmonfhir} served as early proof-of-concept efforts, employing mobile applications and function-calling mechanisms built on Stanford's Spezi ecosystem. While these implementations successfully transformed complex medical data into accessible narratives, they were constrained by platform-specific dependencies, static configurations, and the need for manual prompt engineering. These limitations led to inconsistent outputs and hindered scalability and broader clinical adoption~\cite{teo2023notready}.

Recent work has shifted toward modular, open-source frameworks that prioritize interoperability and reproducibility~\cite{ehtesham2025surveyagentinteroperabilityprotocols}. The integration of the Model Context Protocol (MCP) with FHIR resources represents a significant step in this direction, offering a standardized, declarative interface for accessing diverse healthcare data. Unlike prior approaches that relied on hardcoded API calls, the MCP-FHIR framework enables flexible querying of FHIR resource types, supporting dynamic context injection, transparent reasoning workflows, and more consistent performance in LLM-assisted clinical applications~\cite{mcpfhir, openai_gpt4, llama2}.

These developments reflect a broader movement away from fragmented, custom-built integrations toward unified solutions that emphasize scalability, explainability, and cross-platform interoperability in healthcare AI. By addressing long-standing limitations in context management and structured data access, the proposed framework builds on this progress to advance clinical decision support and improve patient health literacy.

\begin{figure*}[h]
    \centering
    \includegraphics[width=\linewidth]{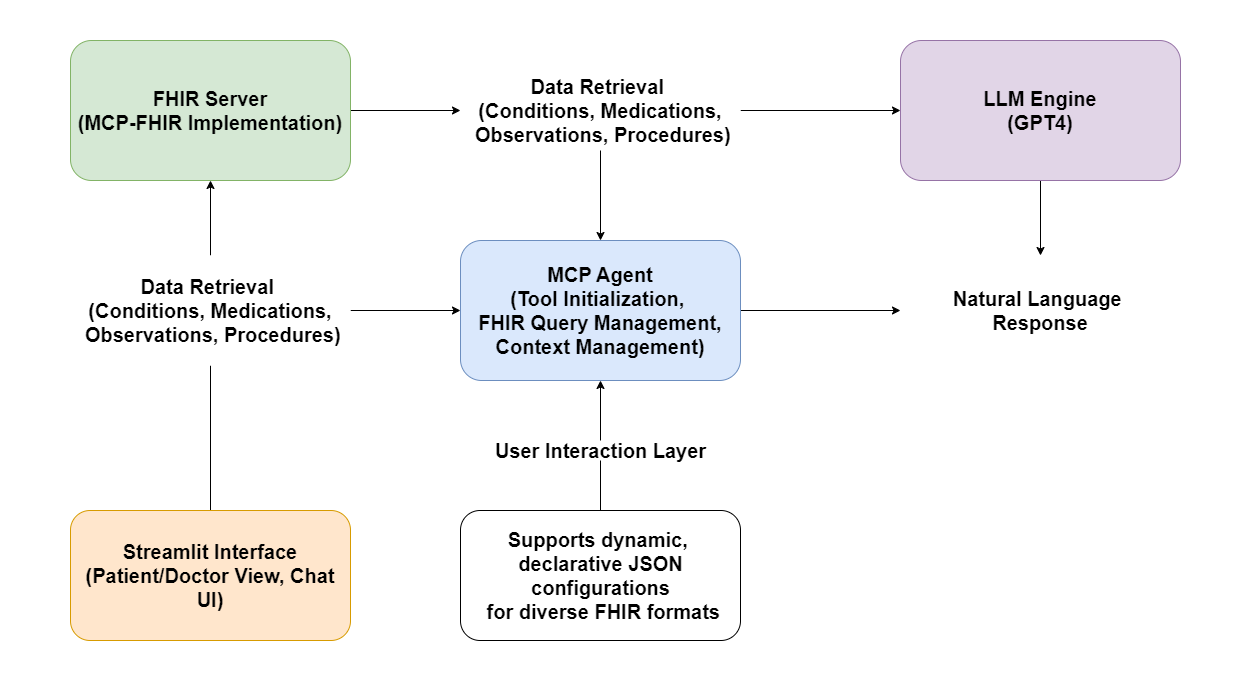}
    \caption{System Architecture using MCP, FHIR, and LLM workflows.}
    \label{fig:architecture}
\end{figure*}

\section{System Overview}
The proposed system addresses persistent challenges in clinical decision support (CDS), documentation overload, and patient health literacy by integrating Large Language Models (LLMs) with FHIR resources using the Model Context Protocol (MCP). This agent-based framework leverages a dynamic, MCP-enabled approach to query patient-specific FHIR data in real time and deliver natural language explanations tailored to various personas, including clinicians, caregivers, and patients.

At the core of the system is an LLM agent that orchestrates secure interactions between structured FHIR resources and the LLM reasoning engine. The integration is facilitated through a lightweight, open-source MCP-FHIR server that supports declarative access to diverse FHIR resource types via JSON configurations. A user-friendly interface, built using the Streamlit framework, allows end users to select patients, review summaries, and engage in conversational interactions with the system.

The operational workflow begins with the initialization of the MCP agent, which establishes connections to the FHIR server and dynamically retrieves patient records, such as conditions, medications, observations, and procedures. Based on the selected patient and persona, the agent composes a context-aware prompt that is transmitted to the LLM. The LLM then produces a synthesized, natural language explanation of the patient data, thereby supporting clinical reasoning and enhancing patient understanding.

This modular and extensible design promotes interoperability, transparency, and reproducibility in clinical workflows. In addition, the architecture is adaptable to various FHIR formats, as exemplified by the MCP-FHIR implementation, and can be extended with additional modules, such as imaging viewers, lab analyzers, and multilingual translation tools.

\section{System Architecture}

The system architecture is designed to support dynamic interactions between LLMs and structured FHIR data via the Model Context Protocol (MCP) . This section details the primary components, workflow, and extensibility features of the framework.

\subsection{Core Components}
The architecture is composed of the following key components:
\begin{itemize}
    \item \textbf{MCP Agent:} Serves as the central orchestrator of clinical workflows by initializing tool connections, selecting appropriate FHIR queries, and managing the context for LLM interactions.
    \item \textbf{FHIR Server:} Based on an open-source MCP-FHIR implementation, this server provides access to FHIR resources. It supports operations such as reading and searching for patient data via declarative JSON configurations.
    \item \textbf{LLM Engine:} Utilizes advanced LLMs (e.g., OpenAI's GPT-4) to synthesize information from FHIR data, delivering concise summaries and interpretations that enhance clinical reasoning and patient understanding.
    \item \textbf{Streamlit Interface:} A lightweight frontend that enables users to select personas (clinician, caregiver, patient), review patient summaries, and interact with the system through conversational queries.
\end{itemize}

\subsection{Workflow Overview}
The system workflow proceeds as follows:
\begin{enumerate}
    \item \textbf{Initialization:} The MCP agent establishes connections with the FHIR server and attaches the LLM engine.
    \item \textbf{Data Retrieval:} Upon selecting a patient and a user persona, the agent queries the FHIR server for relevant EHR data, such as conditions, medications, observations, and procedures.
    \item \textbf{Contextual Prompt Composition:} Based on the user-selected persona and retrieved patient data, the agent composes a tailored, context-aware prompt.
    \item \textbf{LLM Inference:} The composed prompt is transmitted to the LLM engine, which returns a synthesized natural language response that informs clinical decision-making or enhances patient comprehension.
\end{enumerate}

\subsection{Extensibility}
The architecture is inherently modular and extensible:
\begin{itemize}
    \item The MCP framework enables dynamic integration of additional tools and resources, such as imaging viewers or lab analyzers.
    \item The use of declarative JSON configurations allows for seamless expansion to support additional FHIR resource types or alternative data formats.
    \item The composable design encourages reuse and interoperability across various EHR systems and clinical workflows.
\end{itemize}

Figure~\ref{fig:architecture} illustrates the overall architecture, emphasizing the flow between the MCP agent, FHIR server, LLM engine, and user interface.

\section{System Workflow and Implementation}

In this section, the implementation details of the open-source MCP-FHIR clinical assistant system are described. The architecture integrates a configurable Model Context Protocol (MCP) agent, a publicly accessible FHIR server, and an OpenAI LLM through a persona-based Streamlit user interface. This solution supports dynamic querying of FHIR resources and generates contextual clinical summaries in real time, ensuring adaptability, transparency, and applicability in both clinical and patient-facing scenarios.

\subsection{Architecture Overview}

Figure~\ref{fig:architecture} illustrates the overall system architecture. The user interacts with a Streamlit-based front end, where they select a persona (e.g., Clinician, Caregiver, Patient) and choose a patient record for review. The MCP agent functions as an orchestration layer by invoking tools to fetch FHIR resources via standard APIs and then forwarding the structured results to an LLM (e.g., GPT-4o). The LLM generates personalized and comprehensible summaries or responses based on the persona-specific instructions.

\subsection{Agent Configuration with FHIR Server}

The system utilizes the open-source \texttt{mcp-fhir} server (available from Flexpa) which is integrated into the MCP agent through a YAML configuration. This configuration specifies the FHIR base URL and access credentials, thereby enabling the agent to interact with any HL7-compliant resource. The modular configuration approach supports dynamic discovery and retrieval of diverse FHIR resource types in accordance with HL7 standards.

\subsection{Agentic Orchestration and Dynamic Prompt Generation}

The MCP agent is initialized with an instruction set tailored to each persona (for instance, “You are a helpful assistant for a clinician working with EHRs”). The agent leverages standardized tool calls to retrieve clinical data such as Conditions, Medications, Observations, and Procedures from the FHIR server. Each retrieved resource is summarized and subsequently incorporated into a context-aware prompt. For example, a typical prompt may be structured as follows:


\begin{figure}[h]
  \centering
  \begin{chatbox}
    \texttt{Persona: Clinician. \\
    Patient Name: John Doe. \\
    Conditions: Asthma; Hypertension. \\ Medications: Metoprolol; Albuterol.}
  \end{chatbox}
  \label{fig:quote1}
\end{figure}

This prompt, which integrates both demographic and clinical details, is submitted to the LLM through an abstraction layer that manages context history and enforces consistency across interactions. The resulting LLM output is a concise clinical summary or explanatory response intended to support clinical decision-making and enhance patient understanding.

\subsection{Interactive Multimodal User Interface}

The user interface is implemented using Streamlit and is designed to be both intuitive and interactive. Key features include:
\begin{itemize}
  \item A preview of patient lists with demographic summaries.
  \item Persona-specific, predefined question templates to facilitate initial queries.
  \item A free-form chat interface that supports continuous session-aware interactions.
\end{itemize}

Responses are progressively streamed into the UI, enabling real-time conversational interaction with the clinical assistant.

\subsection{Agentic Benefits}

The overall agentic workflow provides several advantages:
\begin{itemize}
  \item \textbf{Modularity:} The system’s architecture permits independent development and reuse of individual tools.
  \item \textbf{Interoperability:} It can connect with any standard FHIR server via the MCP, ensuring broad compatibility.
  \item \textbf{Explainability:} Detailed prompt history and source data are available, supporting transparent clinical reasoning.
  \item \textbf{Scalability:} The architecture is designed for seamless integration with additional modules (e.g., on-device LLMs, clinical validation tools) in future extensions.
\end{itemize}

\vspace{1em}
In summary, the system enables real-time clinical decision support and improved health literacy through a scalable, standards-based architecture. It dynamically supports diverse FHIR data, leverages agentic orchestration for context-aware interactions, and provides interpretable outputs for multiple user personas.

\section{Use Case and Workflow Example}

This section demonstrates a typical end-to-end scenario that highlights the practical value of the MCP-FHIR framework for clinical decision support and patient education. The use case illustrates how dynamic FHIR querying, persona-based prompt generation, and LLM-based summarization are integrated into an interactive workflow.

\begin{figure}[h]
\centering

\includegraphics[width=0.48\textwidth]{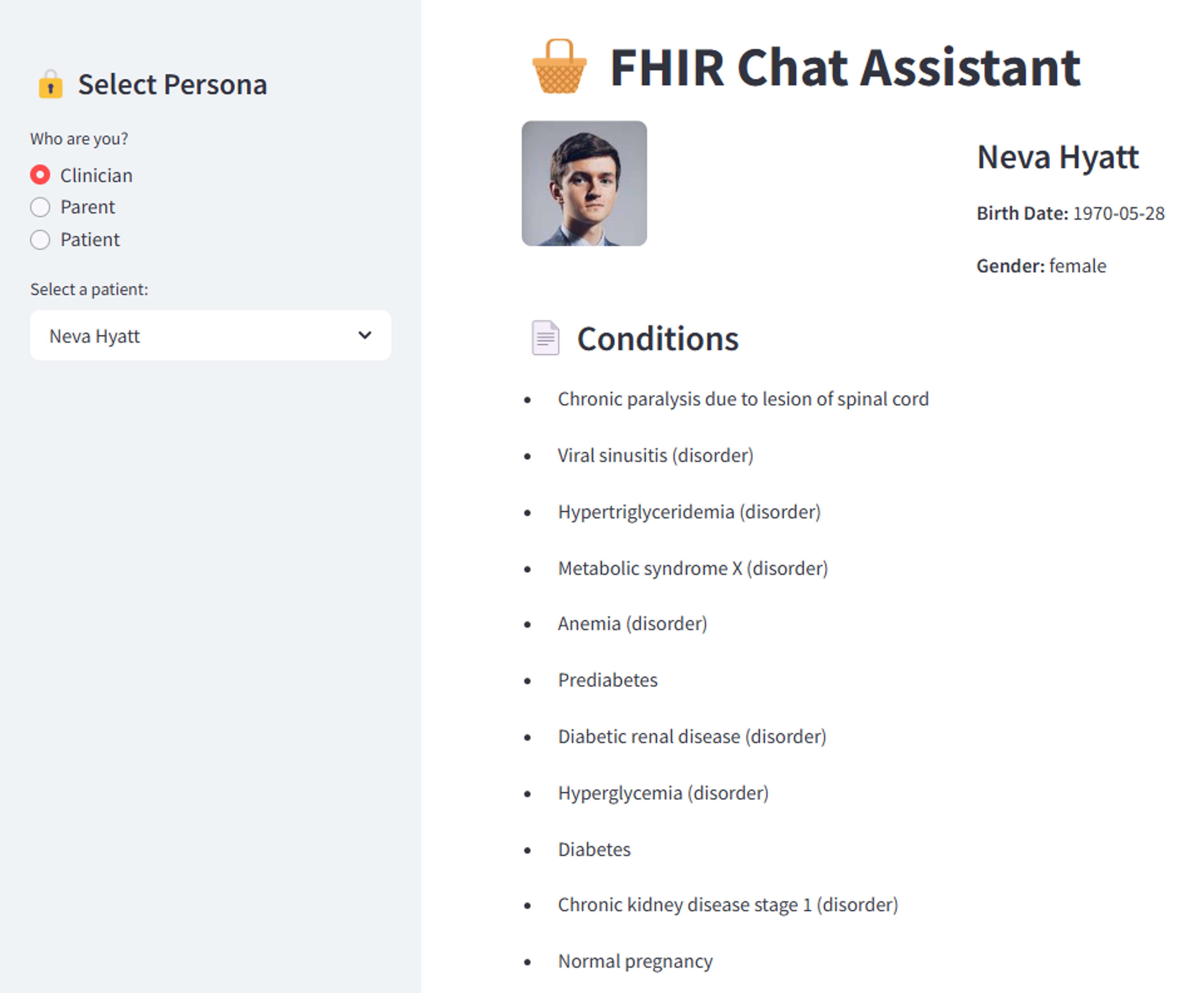}
\caption{An illustration of Clinician Persona.}
\label{fig:usecase}
\end{figure}

\begin{figure}[h]
\centering
\includegraphics[width=0.48\textwidth]{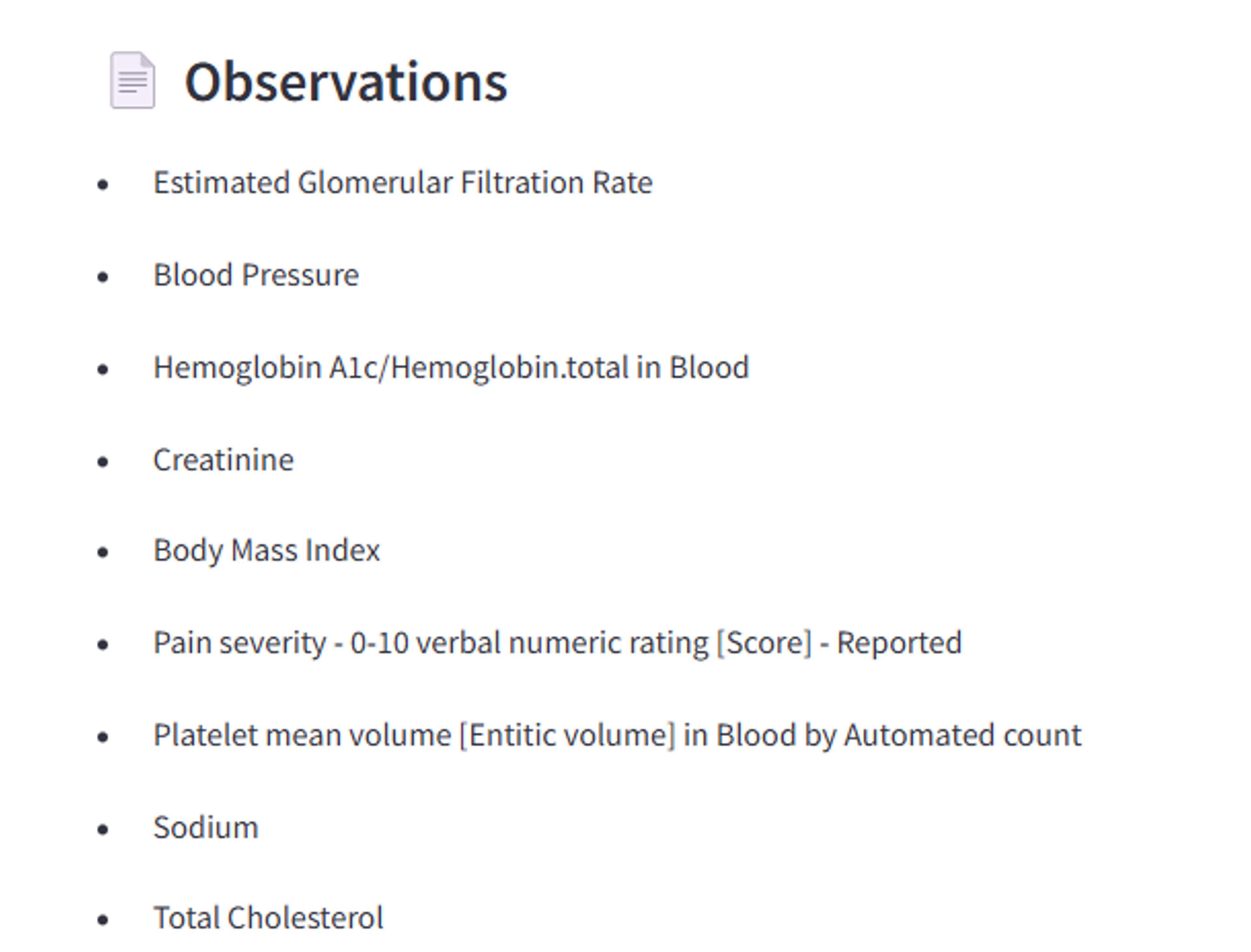}
\caption{An illustration of Observation by Clinician.}
\label{fig:observation}
\end{figure}

\begin{figure}[h]
\centering
\includegraphics[width=0.48\textwidth]{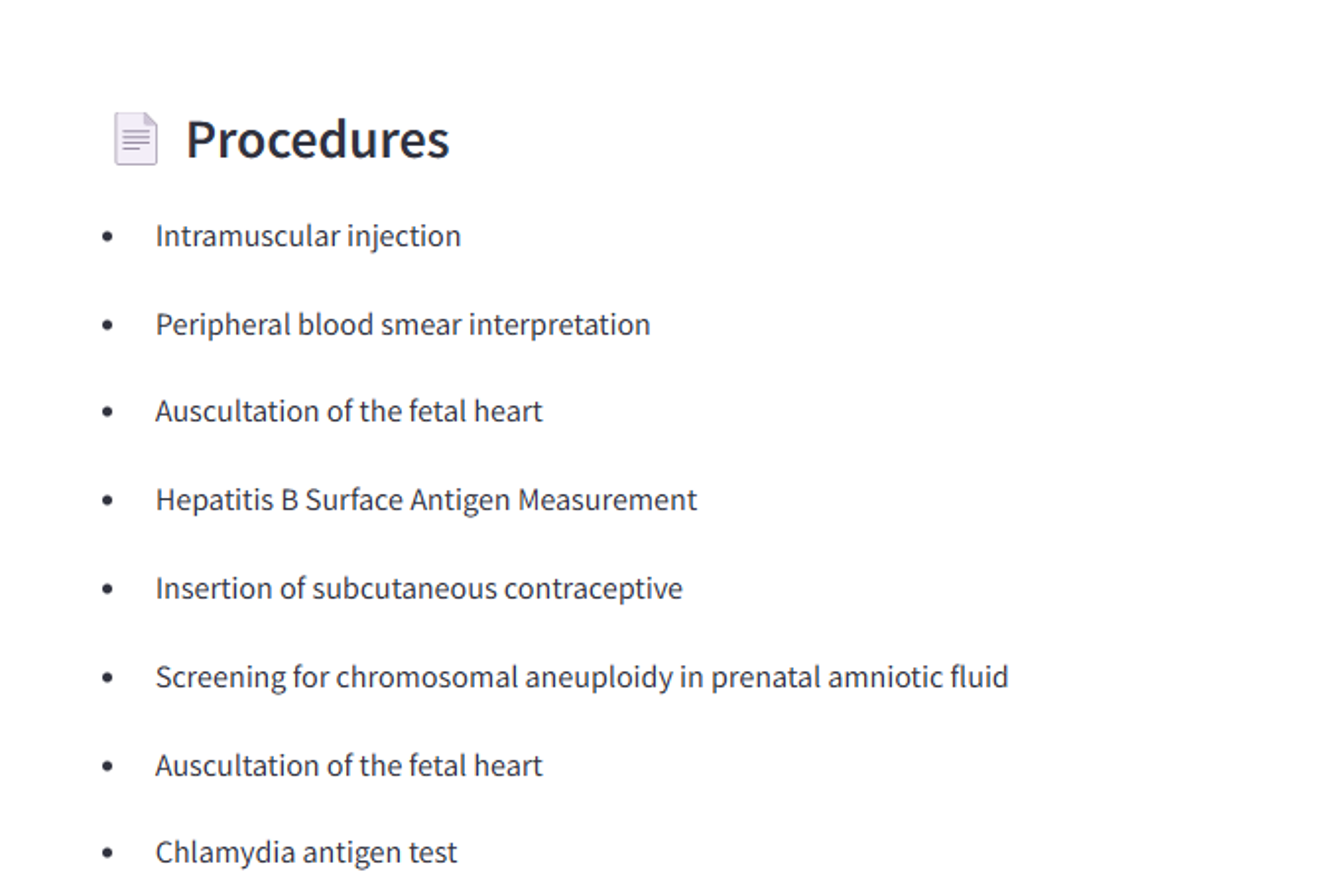}
\caption{A Snippet of Procedures listed by Clinician.}
\label{fig:procedures}
\end{figure}

\begin{figure*}
\centering
\includegraphics[width=\textwidth]{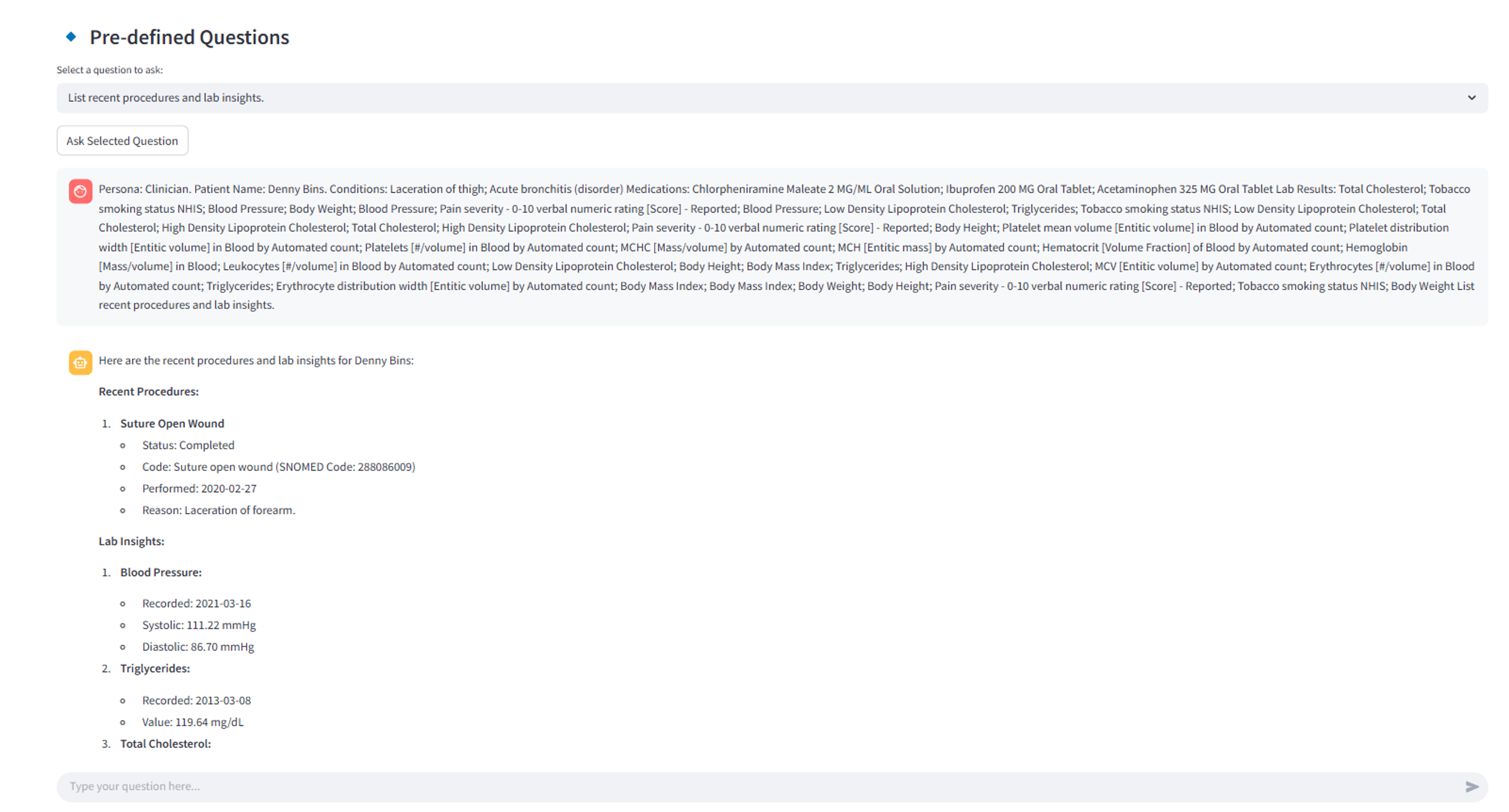}
\caption{An illustration of Pre-defined Questions in a clinician scenario.}
\label{fig:questions}
\end{figure*}

\subsection{Persona-Based Clinical Decision Support Interaction}

In the clinical scenario, a user selects the “Clinician” persona from the Streamlit-based interface. The system then retrieves a list of recent patients, displaying key demographic details such as name, birthdate, and gender. Upon selecting a patient, the MCP agent dynamically fetches multiple FHIR resource types (e.g., \texttt{Condition}, \texttt{MedicationRequest}, \texttt{Observation}, and \texttt{Procedure}). These resources are organized into structured segments, allowing the clinician to quickly gain an overview of the patient's clinical history.

\subsection{Predefined and Freeform Query Options}

To facilitate efficient clinical decision-making, the interface offers predefined query templates, such as:
\begin{itemize}
    \item What treatment options are available for this patient’s conditions?
    \item Summarize the medications and conditions for this patient.
    \item List recent procedures and relevant laboratory insights.
\end{itemize}
When a predefined query is selected, the system composes a context-rich prompt that integrates persona information, patient demographics, and the retrieved FHIR data. The generated prompt is then submitted to the LLM, which returns a concise summary. For example, a typical clinician prompt may yield:

\begin{figure}[H]
  \centering
  \begin{chatbox}
    \textit{Patient John Doe, diagnosed with Type 2 Diabetes and Hypertension, is currently prescribed Metformin and Lisinopril. Recent laboratory results indicate elevated HbA1c levels. Consider revising the treatment regimen and advising lifestyle modifications.}
  \end{chatbox}
  \label{fig:quote2}
\end{figure}


\subsection{Adaptive Workflow for Multiple Personas}

The framework's modular design also supports workflows for other personas, such as caregivers and patients. Under the “Caregiver” persona, the system alters the tone and complexity of the response to ensure the information is accessible. For instance, a caregiver’s query might result in:

\begin{figure}[H]
  \centering
  \begin{chatbox}
    \textit{John Doe has diabetes and high blood pressure. \\
    He is taking medications to control these conditions, and recent tests \\ suggest that his blood sugar levels could be better managed. \\
    It is recommended to review his treatment plan with his physician.}
  \end{chatbox}
  \label{fig:quote}
\end{figure}

This adaptability ensures that clinical insights are communicated effectively according to the user’s role and expertise.

\subsection{Explainability and Session Continuity}

To maintain transparency, each LLM output includes traceable references to the corresponding FHIR resources. This linkage ensures that clinical recommendations can be verified against the original data (e.g., medication names from \texttt{medicationCodeableConcept.text} and laboratory results from \texttt{Observation.code.text}). Moreover, the system retains a session history that preserves context across multiple turns in the conversation, thereby supporting cohesive multi-turn interactions and progressive decision support.

\subsection{Impact on Clinical Workflows}

The described workflow exemplifies several key advantages of the MCP-FHIR solution:
\begin{itemize}
    \item \textbf{Real-Time Data Access:} Dynamic querying eliminates the need for custom API integrations, enabling immediate access to up-to-date FHIR resources.
    \item \textbf{Personalization:} The system adapts the complexity and tone of output to match the selected persona, thereby enhancing usability across clinical and patient settings.
    \item \textbf{Traceability and Explainability:} Each response is directly linked to the underlying FHIR data, ensuring that recommendations are both verifiable and interpretable.
    \item \textbf{Scalability:} The modular architecture supports future extensions, including on-device LLM processing and additional clinical data integrations.
\end{itemize}

In summary, the presented use case demonstrates how the MCP-FHIR framework provides a scalable, standards-based solution for real-time clinical decision support and patient education. The system's ability to dynamically integrate diverse FHIR data, generate context-aware prompts, and produce interpretable outputs underscores its potential to improve clinical workflows and health literacy.

\section{Conclusion}
This paper presented a scalable, standards-based framework that integrates Large Language Models (LLMs) with FHIR resources through the Model Context Protocol (MCP) to enhance clinical decision support and EHR reasoning. The proposed system leverages a modular, agent-based architecture that supports dynamic querying, context-aware prompt generation, and real-time LLM-based summarization. By abstracting the complexities of FHIR data retrieval and integrating persona-specific workflows via a Streamlit interface, the framework offers transparent and reproducible interactions that enhance both clinical decision-making and patient health literacy.

The implementation demonstrates the potential of combining LLMs with standardized FHIR access to generate concise and interpretable clinical insights. Key benefits include the modularity and scalability of the system, the ability to dynamically adjust language output to different user roles, and the provision of traceable, explainable outputs. These advantages are particularly relevant in addressing the growing challenges of documentation overload and the need for personalized health information access.

In summary, the integration of LLMs with MCP-enabled FHIR data access represents a significant step toward more intelligent, interactive, and patient-centered clinical decision support systems.

\bibliographystyle{IEEEtran}
\bibliography{ref.bib}
\end{document}